\documentclass{article}
\usepackage{LaThuileFPSpro}
\usepackage{array}
\usepackage{relsize}
\usepackage{xspace}
\def\babar{\mbox{\slshape B\kern-0.1em{\smaller A}\kern-0.1em
    B\kern-0.1em{\smaller A\kern-0.2em R}}}
\def\pep2{PEP-II}
\def\D0bar{\kern 0.2em\overline{\kern -0.2em D}{\kern 0.1em}\xspace^0}
\def\pep2{PEP-II}
\def\CP{\ensuremath{C\!P}\xspace}
\def\CPV{\ensuremath{C\!PV}\xspace}
\newcommand{\gevc}{\ensuremath{{\mathrm{\,Ge\kern -0.1em V\!/}c}}\xspace}
\newcommand{\mevc}{\ensuremath{{\mathrm{\,Me\kern -0.1em V\!/}c}}\xspace}
\newcommand{\gevcc}{\ensuremath{{\mathrm{\,Ge\kern -0.1em V\!/}c^2}}\xspace}
\newcommand{\mevcc}{\ensuremath{{\mathrm{\,Me\kern -0.1em V\!/}c^2}}\xspace}
\newcommand {\aplt} {\ {\raise-.5ex\hbox{$\buildrel<\over\sim$}}\ }

\begin{document}
%
%
\title{ 
  RESULTS IN CHARM PHYSICS FROM BABAR EXPERIMENT
  }
\author{
  A. Pompili \\ (for the \babar\ Collaboration) \\
  {\em Dipartimento Interateneo di Fisica and INFN,} \\ {\em via Amendola 173, I-70126 Bari, Italy}}
\maketitle
\baselineskip=11.6pt
%
\vspace{-5.5cm} \hspace{8.1cm} SLAC-PUB-10470 
\vspace{5.5cm}
\begin{abstract}
Recent measurements in the charm sector at \babar\ are reviewed. 
The scope of the presentation includes 
the observation of two new narrow mesons in the $D^{+}_{s}\pi^{0}$ and $D^{+}_{s}\pi^{0}\gamma$ final states as well as 
the measurement of $D^0$-$\D0bar$ mixing parameters by means of two methods: using the doubly-Cabibbo-suppressed $D^0$ decay to $K^+\pi^-$ and using the ratios of lifetimes extracted from samples of $D^0$ mesons decaying to $K^-\pi^+$, $K^-K^+$, and $\pi^-\pi^+$.
\end{abstract}
%

\begin{center}
\vspace{5.8cm}
\noindent
Contributed to the Proceedings of the \\ XVIII$^{th}$ Les Rencontres de Physique de la Vall\'{e}e d'Aoste: \\ Results and Perspectives in Particle Physics, \\ 2/29/2004 -- 3/6/2004, La Thuile, Aosta Valley, Italy
\end{center}

\vspace{1.4cm}
\hspace{-1cm}\begin{tabular}{c}
\noindent \textit{Stanford Linear Accelerator Center, Stanford University, Stanford, CA 94039} \\
\hline
\noindent
Work supported in part by Department of Energy contract DE-AC03-76SF00515.
\end{tabular}
\newpage
\baselineskip=14pt
%
\section{Introduction}

B-factories are a powerful tool for charm physics: the effective $e^+e^- \! \! \rightarrow \! c\overline{c} \, (b\overline{b})$ cross section at the energy of the $\Upsilon(4S)$ is 1.30(1.05)${\rm nb}$. The recent \babar\ charm results reviewed in this paper concern $D^0$-$\D0bar$ mixing with 2-body hadronic decays and $c \overline{s}$ spectroscopy.

Mixing searches are discussed (Sec. 2) and results from two different 
experimental methods are presented~\cite{cc}: 
\textit{lifetime difference} searches using $D^0$ Cabibbo-suppressed decays (CSD) 
to $K^-K^+$ and $\pi^-\pi^+$ (Sec. 2.1) and 
\textit{wrong-sign} searches using $D^0$ doubly-Cabibbo-suppressed decay (DCSD) 
to $K^+\pi^-$ (Sec 2.2). The $D^0$ Cabibbo-favoured decay (CFD) to $K^- \pi^+$ is needed in both cases as reference mode.  
The former(latter) analysis is based on a 91(57.1)~${\rm fb}^{-1}$ data sample 
collected on or near the $\Upsilon(4S)$ resonance with the \babar\ detector 
at the \pep2 asymmetric-energy $e^+e^-$ storage ring~\cite{babpep2}. 

The observation of the two narrow states $D^*_{sJ}(2317)$ and $D_{sJ}(2458)$
(Sec. 3) is based on an integrated luminosity of 91${\rm fb}^{-1}$.

The kinematical selection from continuum is obtained by requiring a minimum value of about 2.5~\gevc for the momentum of a charmed meson in the center-of-mass frame. This criterium provides full 
removal of charmed mesons from $B$ decays and a strong reduction of the combinatorial background. Enhanced rejection can be obtained by increasing this threshold when needed as in the spectroscopic studies of Sec.3. 

The $D^{0} \! \! \rightarrow \! h^{+}h^{-}$ candidates selected in the mixing analyses presented in Sec.2 are mostly $D^{*}$-tagged: they are required to be produced by the decay $D^{*+} \! \! \rightarrow \! D^{0}\pi^{+}_{s}$. The slow pion charge provides the flavour tagging; it also allows to introduce the quantity $\delta m = m(h^+ h^- \pi_{s}) - m(h^+ h^-)$ that is crucial to improve the background rejection.

Vertexing, tracking and particle identification performances are crucial for the analyses presented. The standard kaon identification provided mainly by the Cherenkov detector is characterized  by an average efficiency of about 85\% and an average pion contamination probability of about 2\%.

\section{$D^0$-$\D0bar$ mixing}

The two $D^0$ physical states can be represented as $|D_{1,2} \rangle = p | D^0 \rangle \pm q | \D0bar \rangle$ (with masses $m_{1,2}$ and widths $\Gamma_{1,2}$) where $\left|p\right|^2 + \left|q\right|^2 = 1$ and $| D^0 \rangle$,$| \D0bar \rangle$ are the flavour eigenstates. The mixing parameters $x \equiv m_{1}-m_{2}/\Gamma$, $y \equiv \Gamma_{1}-\Gamma_{2}/2\Gamma$, where $\Gamma$ is the average width, are predicted by the Standard Model (SM) to be very small in the charm sector: $|x|,|y|\le 10^{-3}$; however new physics contributions may enhance $x$, whereas final state interactions and SU(3) breaking can enhance $y$~\cite{burdship}. Thus charm mixing provides a sensitive probe for physics beyond the SM at the level of the current experimental sensitivity ($\sim \! 10^{-2}$). 

Since new physics may not conserve \CP, it is necessary to consider \CP violation (\CPV) when measuring mixing. \CPV effects can be parametrized in terms of the following quantities ($D^0$,$\D0bar$ decay into a final state $f$):
\begin{equation}
r_{m} \equiv \biggl | \frac{q}{p} \biggr | \, , \, \varphi_f \equiv arg \biggl ( \frac{q}{p} \frac{\overline{A}_{f}}{A_{f}}\biggr ) \, ; \, A_f \equiv \langle f | {\mathcal H}_D | D^0 \rangle \, , \, \overline{A}_f \equiv \langle f | {\mathcal H}_D | \D0bar \rangle
\label{cpvparam} 
\end{equation}
A value of $r_m \neq 1$ would indicate \CPV in mixing; 
a value of $\varphi_f \neq 0$ would indicate \CPV in the interference of mixing and decay. 
%
%
\subsection{Lifetime difference searches}

Mixing would alter the decay time distributions of $D^0$ and $\D0bar$ mesons that decay into \CP eigenstates. They can be considered to a good approximation, as pure exponential with effective lifetimes~\cite{bergmann}
\begin{equation}
\tau^\pm = \tau^0\left[ 1 + r_m^{\pm1} \left( y\cos\varphi_f \mp
                                 x\sin\varphi_f \right) \right]^{-1} 
\end{equation}
where $\tau^0$ is the lifetime for the CFD $D^0 \to K^-\pi^+$ (and $\D0bar \to K^+\pi^-$) and $\tau^+$ ($\tau^-$) is the lifetime for the CSD of the $D^0$ ($\D0bar$) into \CP-even final states (such as $K^-K^+$ and $\pi^-\pi^+$).
They can be combined to build the parameters 
\begin{equation}
Y = \frac{\tau^0}{\langle\tau\rangle} - 1 \, , \;
\Delta Y = \frac{\tau^0}{\langle\tau\rangle} A_\tau \; , \; \;
\mbox{where} \, \left\{ \begin{array}{ll} 
\! \langle \tau \rangle = (\tau^++\tau^-)/2 \\
\! A_\tau = (\tau^+-\tau^-)/(\tau^++\tau^-)
\end{array} \right.
\label{yparam}
\end{equation}
Both $Y$ and $\Delta Y$ are zero if there is no mixing. 
Otherwise, in the limit of \CP conservation in mixing,
$Y = y\cos\varphi_f$, $\Delta Y = x\sin\varphi_f$; 
moreover, in the \CP conserving limit in the interference between mixing and decay,
$Y=y$, $\Delta Y = 0$. In these limits $Y$ is not a new physics search parameter but simply a measure of the mixing level due to SM.

\begin{figure}[t!]
\includegraphics{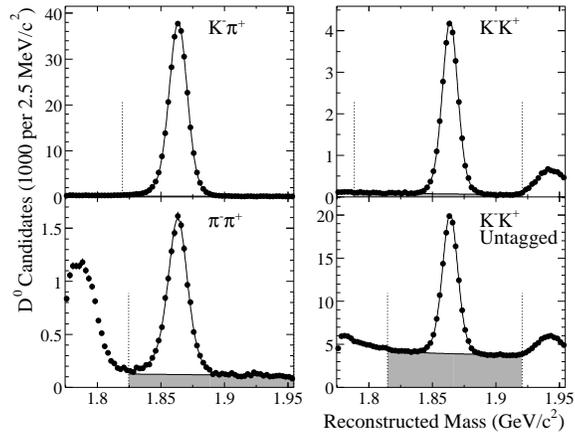}
\vspace{5.3cm}
\caption{\it The reconstructed $D^0$ invariant mass distributions (points) and the fit result (superimposed curve) for the four samples. The portion of each sample assigned by the fit to the background is indicated by the shaded region.}
\label{fig:ymass} 
\end{figure}

\begin{figure}[b!]
\includegraphics{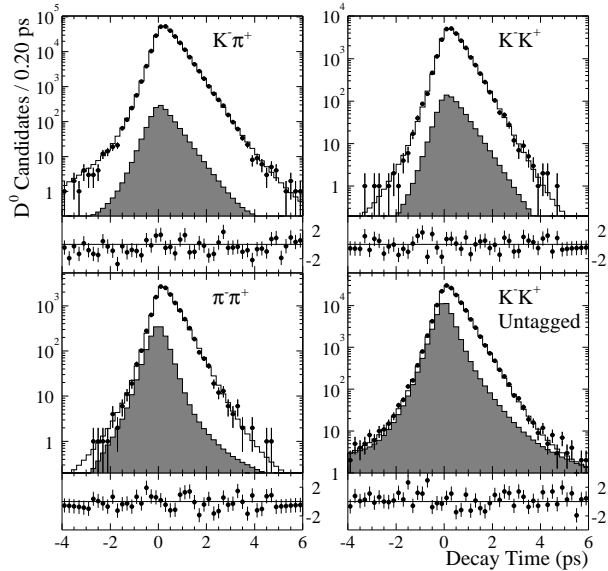}
\vspace{7.4cm}
\caption{\it The $D^0$ proper time distribution within a $\pm 15~MeV/c^2$ mass signal window (points) and the fit result (superimposed histogram) for the four samples. The shaded histogram is the portion assigned by the fit to the background. The points beneath each plot show the bin-by-bin difference between data and fit divided by the statistical error.}
\label{fig:yfit} 
\end{figure}

The $D^0$ mass distribution for the four independent selected samples are shown in fig.\ref{fig:ymass}. Three of them are $D^*$-tagged: $K^-\pi^+$ measures $\tau^0$, $K^-K^+$ and $\pi^-\pi^+$ both measure $\langle \tau \rangle$ and $A_\tau$. The untagged sample $K^-K^+$, about six times wider than the tagged one at the cost of a purity drop from 97\% to 68\%, measures only $\langle \tau \rangle$. The mass fits determine the probability of each selected $D^0$ to belong to a signal event that is used in the proper time fit. Sidebands candidates are included in the fit sample to better constrain the background. The lifetime for each sample is extracted by an unbinned maximum likelihood (UML) fit; the likelihood is a sum of two probability density functions (PDFs), one for the signal and one for the background, each differently modelled as the convolution of a decay-time distribution and a resolution function. The fit results are shown in fig.\ref{fig:yfit}. The average $\langle \tau \rangle$ and the asymmetry $A_\tau$ are calculated from  the lifetimes $\tau^0$ and $\tau^\pm$. Systematics effects on lifetime tend to cancel in the lifetime ratio. This analysis is described in detail elsewhere~\cite{ybabar}.

The results, summarized in tab.\ref{tab:ydyresults}, are consistent with the absence of mixing and \CPV within the current sensitivity. A comparison with other experiments is given in tab.\ref{tab:ycompare}. \babar\ provides the most stringent limits on $Y$ and the first measurement of a \CPV parameter with the lifetime ratio method.

The $Y$ measurement is statistically dominated (by CSD) and a 1\% sensitivity, for a single tagged sample, is within the reach of the B-factories. 

\begin{table}[t]
\centering
\caption{ \it $Y$ and $\Delta Y$ results; the first error is statistical, the second systematic.}
\vskip 0.06 in
\begin{tabular}{|l|l|l|}
\hline
\hspace{1cm} Sample & \hspace{0.7cm} $Y$ (\%) & \hspace{0.7cm} $\Delta Y$ (\%) \\
\hline \hline
$K^-K^+$ & $1.5 \pm 0.8 \pm 0.5$ & $-1.3 \pm 0.8 \pm 0.2$ \\
$\pi^-\pi^+$ & $1.7 \pm 1.2 \mbox{}^{ \>+\,1.2}_{ \>-\,0.6} $ & \hspace{0.15cm} $0.3 \pm 1.1 \pm 0.2$ \\
$D^{*}$-untagged $K^-K^+$ & $0.2 \pm 0.5 \mbox{}^{ \>+\,0.5 }_{ \>-\,0.4}$ & \hspace{1cm} --- \\
\hline
\hspace{0.8cm} Combined & $0.8 \pm 0.4 \mbox{}^{ \>+\,0.5 }_{ \>-\,0.4}$ & $-0.8 \pm 0.6 \pm 0.2$ \\ \hline 
\end{tabular}
\label{tab:ydyresults}
\end{table}

\begin{table}[b]
\centering
\vspace{-0.3cm} \caption{\it Comparison for $Y$ and \CPV results (Belle's $A_{\Gamma}$ is $A_\tau$ of eq.~\ref{yparam}).}
\vskip 0.06 in
\begin{tabular}{|l|l|c|}
\hline
Experiment & \hspace{0.75cm} $Y$ (\%) & \CPV parameter (\%) \\ \hline \hline
\babar\ combined~\cite{ybabar} & \hspace{0.15cm} $0.8 \pm 0.4 \mbox{}^{ \>+\,0.5 }_{ \>-\,0.4 }$ &  $\Delta Y = -0.8 \pm 0.6 \pm 0.2$ \\ \hline
Focus~\cite{yfocus} & \hspace{0.15cm} $3.4 \pm 1.4 \pm 0.7$ & --- \\ \hline 
CLEO II~\cite{ycleo} & $-1.2 \pm 2.5 \pm 1.4$ & --- \\ \hline \hline
Belle $K^-K^+$ (\textit{prelim.})~\cite{ybelle} & \hspace{0.15cm} $1.2 \pm 0.7 \pm 0.4$ & $A_{\Gamma} = -0.2 \pm 0.6 \pm 0.3$ \\ \hline
Belle $D^{*}$-untagged $K^-K^+$~\cite{ybelleuntagged} & $-0.5 \pm 1.0 \mbox{}^{ \>+\,0.7 }_{ \>-\,0.8}$ & --- \\ \hline
\end{tabular}
\label{tab:ycompare}
\end{table}

\subsection{Wrong sign searches in hadronic decays}

This method implies the study of the time evolution of the DCSD $D^0 \! \rightarrow \! K^+\pi^-$, the so called wrong-sign(WS) decay, looking for a mixing signal followed by a CFD: $D^0 \! \rightarrow \! \D0bar \! \rightarrow \! K^+ \pi^-$. 
Since DCSD and CFD may be characterized by different final state interactions, there is an unknown relative strong phase between their amplitudes, $\delta_{K\pi}$, not measurable in a mixing analysis. 
Assuming that $x,y \! \ll \! 1$ and \CP is conserved ($r_{m} \! =\! 1$, $\phi \! = \! 0$), the time-dependent WS decay rate is given by
\begin{equation}
\Gamma_{WS} (t) \propto e^{\Gamma t} \Bigl [ R_{D} + y' \sqrt{R_{D}} \Gamma t + \frac{1}{4}(x'^2 + y'^2)\Gamma^{2}t^{2} \Bigr ]  \; \; , \; \; \Gamma = \frac{1}{\tau^{0}} 
\label{gws} 
\end{equation}
\begin{equation}
\mbox{where} \; \; \; R_{D}=\frac{\Gamma(D^0 \! \rightarrow \! K^+ \pi^-)}{\Gamma(\D0bar \! \rightarrow \! K^+ \pi^-)} \; , \; \left\{ \begin{array}{ll} x'= \; \; \; xcos\delta_{K\pi} + ysin\delta_{K\pi} \\ y'=-xsin\delta_{K\pi} + ycos\delta_{K\pi} \end{array} \right.
\end{equation}
The second and third terms in eq.~\ref{gws} introduce a deviation from a pure exponential; the third one is the mixing term whereas the second one is the interference term that enhances the mixing signal. The time-integrated WS decay rate is
\begin{equation}
R_{WS} = R_{D} + y' \sqrt{R_{D}} + R_{M} \; , \; R_{M} = (x'^2 +y'^2)/2
\end{equation}
where $R_{M}$ is the mixing rate. Additional \CPV effects are included by applying eq.\ref{gws} to $D^0$ and $\D0bar$ separately: $\{ R_{D},x'^2,y' \} \! \rightarrow \! \{ R^{\pm}_{D} , x'^{\pm ^{2}}, y'^{\pm} \}$. The quantities $A_{D}$, $A_{M}$ are related to \CPV in the DCSD and mixing respectively, whereas \CPV in the interference of DCSD and mixing is parametrized by the phase $\phi$:
\begin{equation}
A_{D,M} \! =\frac{R^{+}_{D,M}\! -R^{-}_{D,M}}{R^{+}_{D,M} \! +R^{-}_{D,M}} \; ; \left\{ \begin{array}{ll} \! \! x'^{\pm} \! = C_{M} (x'cos\phi \pm y' sin\phi) \\ \! \! y'^{\pm} \! = C_{M} (y'cos\phi \pm x' sin\phi) \end{array} \right. \! \! \! \! , \, C_{M} \! =\sqrt[4]{\frac{1\pm A_{M}}{1\mp A_{M}}}
\end{equation}

Each reconstructed candidate is assigned to one of the four categories based on its origin ($D^0$ or $\D0bar$) and its decay (RS or WS). The mixing parameters are determined by an extended UML fit to the RS and WS samples simultaneously in a four-dimensional space of variables: mass $m_{K\pi}$, mass difference $\delta m$, proper time $t$ and its error $\sigma_{t}$. Within a category, the likelihood is a sum of PDFs, one for each signal or background component, weighted by the number of events for that component; each component's PDF is modelled as the convolution of a decay-time distribution and a resolution function.  

\begin{figure}[t!]
\includegraphics{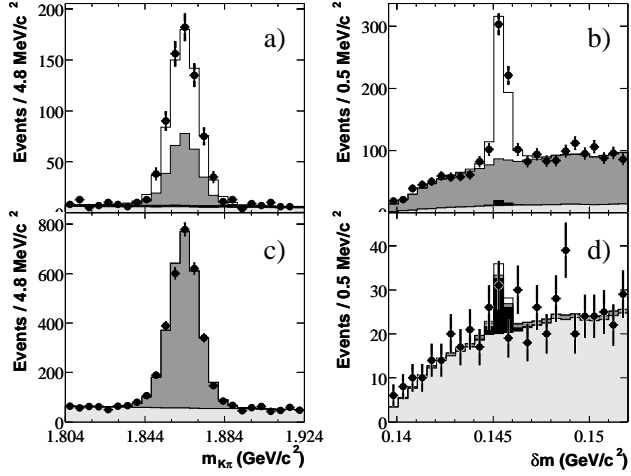}
\vspace{5.85cm}
\caption{\textit{Wrong-sign distributions of \textbf{(a,c)} $K^+\pi^-$ invariant mass and \textbf{(b,d)} $\delta m$ mass difference, for \textbf{(a,b)} signal regions and \textbf{(c,d)} sidebands regions. Data are shown as points with the fit contributions superimposed: signal (open histogram), unassociated slow pion background (dark shaded), double misidentification background (black) and combinatorial background (light shaded).}}
\label{fig:wrongmass} 
\end{figure}

The fit is performed in two main steps. Firstly a fit is done in the $m_{K\pi}-\delta m$ plane to extract the number of candidates in the signal and background components. Its result is shown in fig.~\ref{fig:wrongmass} for the WS sample; the high level background has two peaking components: one for $m_{K\pi}$, associated to a real $D^0$ with wrong slow pion, and one for $\delta m$, associated to $K\pi$ doubly misidentified. There are about 440 WS signal decays. Secondly, a simultaneous fit to the RS and WS decay time distributions is performed. The larger and clean RS sample fixes $\tau_0$ and the resolution model parameters for unmixed decays, while the mixing parameters are determined only by the WS sample. The background time distributions are determined from $m_{K\pi}$ and $\delta m$ sidebands in data. 

\begin{table}[t]
\centering
\vspace{-0.3cm} \caption{\it Wrong-sign results: $95\%$ C.L. intervals including systematic errors.}
\vskip 0.06 in
\begin{tabular}{|c||c|c|c|c|}
\hline
$~$ & \multicolumn{4}{|c|}{$95\%$ C.L. range or value (all $\times 10^{-3}$)} \\ \cline{2-5}
$~$ & CPV allowed & No CPV & No Mixing & No CPV nor Mixing \\ \hline \hline
$R_{D}$ & $(2.3,5.2)$ & $(2.4,4.9)$ & $3.57 \pm 0.22 \pm 0.27$ & $3.59 \pm 0.20 \pm 0.27$ \\ \hline
$A_{D}$ & $(-2.8,4.9)$ & --- & $95 \pm 61 \pm 83$ & ---  \\ \hline
$x'^{2}$ & $< 2.2$ & $< 2.0$ & --- & --- \\ \hline
$y'$ & $(-56,39)$ & $(-27,22)$ & --- & --- \\ \hline
$R_{m}$ & $< 1.6$ & $< 1.3$ & --- & --- \\ \hline
\end{tabular}
\label{tab:wsresults}
\end{table}

\begin{figure}[b!]
\includegraphics{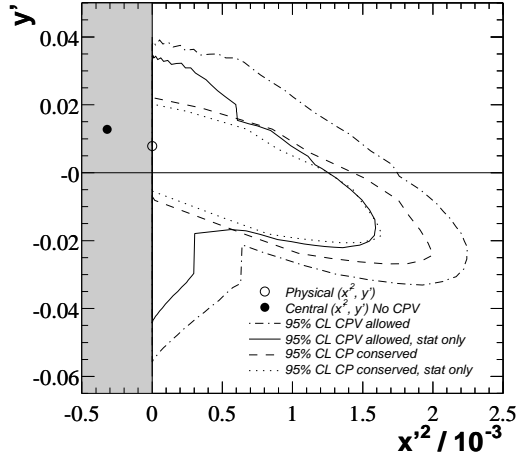}
\vspace{6.31cm}
\caption{\it Fit results and $95\%$ C.L. contours in the $(x'^2 ,y)$ plane. The solid point represents the most likely fit result assuming \CP conservation and the open circle the same but allowing \CPV and forcing $x'^2 > 0$. The dotted (solid) line is the statistical contour for \CP conservation (\CPV allowed); the dashed (dash-dotted) line is the corresponding contour when systematics are added.}
\label{fig:wrongfits} 
\end{figure}

A procedure based on toy Monte Carlo experiments is used to set frequentistic 95\% C.L. contours in the fitting parameters (shown in fig.~\ref{fig:wrongfits}). This analysis is described in detail elsewhere~\cite{wsbabar}. The results in tab.\ref{tab:wsresults} are consistent with the absence of mixing and \CPV.

\begin{table}[b]
\centering
\vspace{-0.3cm} \caption{\textit{Partial (no \CPV) comparison of results of the wrong-sign analyses.}}
\vskip 0.06 in
\begin{tabular}{|l||l||c|c|} \hline
Experiment & \hspace{0.65cm} $R_{WS}$ (\%) & $x'^{2}$(\%) & $y'$(\%) \\ \hline \hline
\babar\ ~\cite{wsbabar} & $0.357 \pm 0.022 \pm 0.027$ & $< 0.200$ & $(-2.7,2.2)$ \\ \hline
Focus (\textit{prelim.})~\cite{wsfocus} & $0.404 \pm 0.085 \pm 0.025$ & $< 0.152$ & $(-12.4,-0.6)$ \\ \hline
CLEO II~\cite{wscleo} & $0.332 \mbox{}^{ \>+\,0.063 }_{ \>-\,0.065} \pm 0.040$ & $< 0.076$ & $(-5.2,0.2)$ \\ \hline
Belle (\textit{prelim.})~\cite{wsbelle} & $0.372 \pm 0.025 \mbox{}^{ \>+\,0.009 }_{ \>-\,0.014 }$ & --- & --- \\ \hline
\end{tabular}
\label{tab:wscompare}
\end{table}

A comparison of $R_{WS}$, $x'^2$ and $y'$ among experiments is given in tab.\ref{tab:wscompare}. Assuming no mixing, the WS time integrated decay rate is just the rate of the DCSD. For mixing parameters results, a direct comparison is possible only when \CP conservation is assumed. \babar's limits include systematic uncertainties and have been obtained allowing $x'^2 \! <0$. 
The rather different techniques used to obtain limits and treatment of the fit parameter $x'$ may explain the reason for which, in spite of a data sample six times larger, the \babar\ upper limit for $x'^2$ is almost three times larger than CLEO's (two times when considering the systematic error $\pm 0.04\%$ quoted by CLEO). 
Moreover toy Monte Carlo studies~\cite{ysign} show that for a given experiment, when the preferred sign of $y'$ is positive(negative) as in \babar(CLEO) analysis, the allowed region becomes larger(smaller). Naively this can be explained by a partial cancellation (that takes place with $y' \! <0$) between the interference and mixing terms in eq.~\ref{gws} so that the fit becomes sensitive to small changes of the mixing parameters.

\hspace{-0.18cm} At B-factories the WS analyses can improve $x$ sensitivity towards 1\% level. Promising techniques have been proposed to measure the relative strong phase.

\section{Two new charmed mesons: $D^*_{sJ}(2317)$ and $D_{sJ}(2458)$}

\babar\ has observed two new narrow states in the $D^{+}_{s}\pi^{0}$~\cite{babar2317} and $D^{+}_{s}\pi^{0}\gamma$~\cite{babar2458} invariant mass spectra, from inclusive $e^{+}e^{-}$ interactions, denoted as $D^{*}_{sJ}(2317)^{+}$ and $D_{sJ}(2458)^{+}$ respectively. Both states have been observed by CLEO~\cite{dsjcleo} and Belle~\cite{dsjbelle} experiments as well. Belle~\cite{dsjbellefromb} and recently \babar~\cite{moriond04} have reported evidence for both states not only in continuum but also in $B$ decays.

\begin{figure}[t!]
\begin{tabular}{cc}
\includegraphics{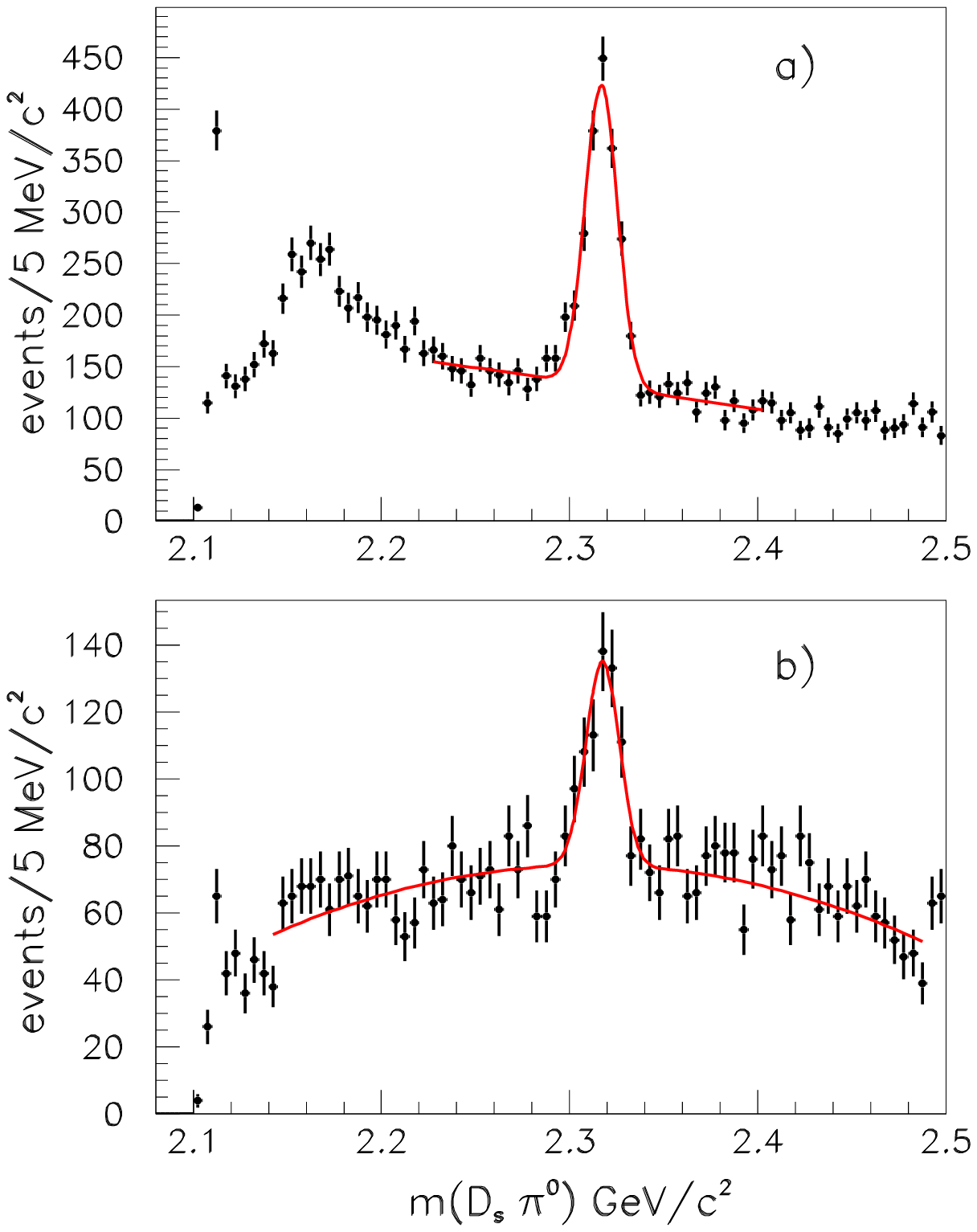} \\ 
\includegraphics{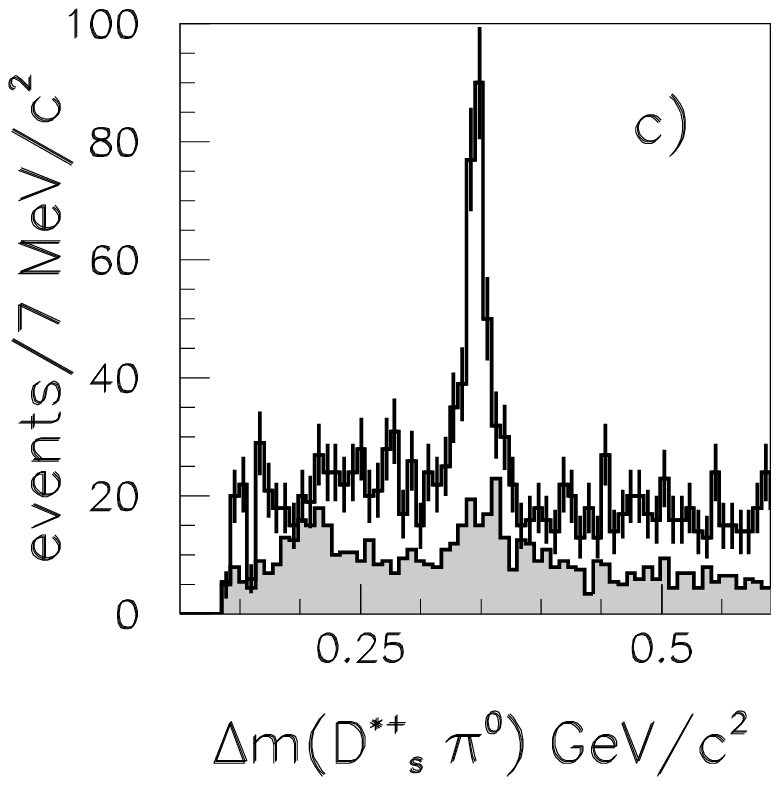}
\end{tabular}
\vspace{4.45cm}
\caption{\it The $D^{+}_{s}\pi^{0}$ mass distribution for the decays \textbf{(a)} $D^{+}_{s} \! \rightarrow \! K^{+}K^{-}\pi^{+}$ and \textbf{(b)} $D^{+}_{s} \! \rightarrow \! K^{+}K^{-}\pi^{+}\pi^{0}$; the fit results (curves) for the $D^{*}_{sJ}(2317)^{+}$ signal are superimposed. The very narrow peak near threshold is associated to the $D^{*+}_{s} \! \rightarrow \! D^{+}_{s} \pi^{0}$ decay. \textbf{(c)} The $\Delta m_{\pi^{0}}$ distribution for $D^{+}_{s}\pi^{0}\gamma$ combinations in the $D^{*+}_{s}$ signal region (points) and sidebands (shaded histogram).}
\label{fig:2317-2458} 
\end{figure}

\begin{figure}[b!]
\includegraphics{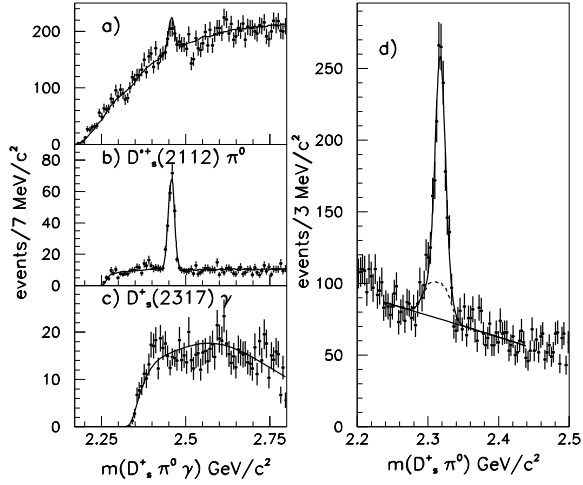}
\vspace{6.37cm}
\caption{\it Fit results from channel likelihood method (curves) overlaid on the $D^{+}_{s} \pi^{0} \gamma$ mass distribution with \textbf{(a)} no weights and after applying weights corresponding to the decays to \textbf{(b)} $D^{*+}_{s}\pi^{0}$ and \textbf{(c)} $D^{*}_{sJ}(2317)^{+} \gamma$. \textbf{(d)} The $D^{+}_{s}\pi^{0}$ mass spectrum (with no additional $\gamma$ requirement). The fit results are superimposed; the dashed and lower solid curves show the contributions from $D_{sJ}(2458)^{+}$ decays and combinatorial background respectively.}
\label{fig:maxlik} 
\end{figure}

The observation for the $D^{*}_{sJ}(2317) \! \rightarrow \! D^{+}_{s}\pi^{0}$ is given in figs.~\ref{fig:2317-2458}a and~\ref{fig:2317-2458}b for $p^{*}(D^{+}_{s}\pi^{0}) \! > \! 3.5$~\gevc and two different $D^{+}_{s}$ decay modes. 
The observation for the $D_{sJ}(2458)^{+} \! \rightarrow \! D^{+}_{s}\pi^{0}\gamma$ is provided by the peak in the distribution of the quantity $\Delta m (D^{*+}_{s}\pi^{0}) \! \equiv \! m(D^{+}_{s}\pi^{0}\gamma) \! - \! m(D^{+}_{s}\pi^{0})$ shown in fig.~\ref{fig:2317-2458}c: the peak is missing when considering $D^{*}_{s}$ sidebands; the residual background would correspond to $D^{*}_{sJ}(2317)$ mesons associated to random photons selected to be consistent with the $D^{*}_{s} \! \rightarrow \! D_{s} \gamma$ decay. 
To disentangle these two possible decays and evaluate a possible $D^{*}_{sJ}(2317)$ feed-up effect, an UML fit using the channel likelihood method is applied; this fit is simultaneously performed to the $D_{s}\pi^{0}\gamma$, $D_{s}\pi^{0}$ and $D_{s}\gamma$ mass spectra of all $D_{s}\pi^{0}\gamma$ combinations. 
Fit results are shown in the three plots at left-side of fig.~\ref{fig:maxlik}. No clear indication of decays to $D^{*}_{sJ}(2317)\gamma$ is found (the estimated limit is given in tab.\ref{tab:dsj}). 

On the other hand a feed-down effect due to $D_{sJ}(2458)$ has been investigated; it would be produced as follows: $D_{sJ}(2458)^{+} \! \rightarrow \! D^{*+}_{s} \pi^{0}$ and $D^{*+}_{s} \! \rightarrow \! D^{+}_{s} \gamma$ where the $D^{+}_{s} \pi^{0}$ combination could emulate a $D^{*}_{sJ}(2317)^{+}$ candidate and the residual photon would appear to be unrelated or simply lost. Fig.~\ref{fig:maxlik}d shows the fitted $D_{s}\pi^{0}$ mass spectrum in data with the peaking background contribution, estimated from Monte Carlo simulation, associated to $D_{sJ}(2458)$ decays. 

From previous fits, masses (given in tab.\ref{tab:dsj}), widths and yields are extracted; from these yield, once corrected by recostruction efficiency, the relative production rate in continuum is calculated (tab.\ref{tab:dsj}). The observed widths ($\aplt \!$ 10~Me\kern -0.1emV) are consistent with the experimental resolution as determined by simulation studies; thus the intrinsic width must be below the observed one. 
This analysis is described in detail elsewhere~\cite{babar2317}~\cite{babar2458}.

Further experimental information on the decay modes of $D^{*}_{sJ}(2317)$ and $D_{sJ}(2458)$ (including those produced in $B$ decays) allows one to infer their most likely spin-parity assignment~\cite{qcdatwork}: $0^{+}$ and $1^{+}$ respectively. Therefore they could be identified with the missing $j \! = \! 1/2$ doublet of P-wave $c\overline{s}$ states. 
Their masses are significantly lower than predicted by established potential models (below $DK$ and $D^{*}K$ thresholds respectively) and enough to close off the most natural decay modes. If they are interpreted as ordinary $c \overline{s}$ states, this forces them to decay mainly via isospin-violating transition making their widths quite narrow; a possible mechanism is the decay through a virtual $\eta$ followed by $\eta-\pi^{0}$ mixing. 
The unexpected masses and widths have led to a great effort on theoretical side as well~\cite{thbiblio}: either these models need modification or the observed state might have a non-$q\overline{q}$ nature. Among the ``exotic'' explanations, a $DK$ molecule, a four-quark state and a $D\pi$ atom have been proposed; among ``ordinary'' explanations there have been attempts to revise or modify potential models and to couple chiral perturbation theory with HQET; it has been even proposed that these states are a mixture of $c\overline{s}$ and four-quark states.

Further experimental information on the radiative decays and di-pion emission of the new mesons can definitely shed light on their nature. 

\begin{table}[b]
\centering
\vspace{-0.5cm} \caption{\textit{Summary of $D^{*}_{sJ}(2317)$ and $D_{sJ}(2458)$; masses are in \mevcc.}}
\vskip 0.06 in
\setlength{\extrarowheight}{0.06cm}
\begin{tabular}{|@{\hskip 0.06cm}l@{\hskip 0.06cm}|@{\hskip 0.06cm}c@{\hskip 0.06cm}|@{\hskip 0.06cm}c@{\hskip 0.06cm}|@{\hskip 0.06cm}c@{\hskip 0.06cm}@{\hskip 0.06cm}|} \hline
\small Measured quantity & \babar~\cite{babar2317}~\cite{babar2458} & Belle(cont.)~\cite{dsjbelle} & CLEO~\cite{dsjcleo} \\ \hline \hline
\footnotesize $D^+_{s}$ modes & $\phi \pi^+,\overline{K}^{*0}K^+$ & $\phi \pi^+$ & $\phi \pi^+$ \\ \hline 
\footnotesize $m(D^{*}_{sJ}(2317))$ & 
\footnotesize $2317.3 \pm 0.4 \pm 0.8$ & 
\footnotesize $2317.2\pm 0.5 \pm 0.9$ & 
\footnotesize $2318.5 \pm 1.2 \pm 1.1$ \\ \hline
\footnotesize $m(D_{sJ}(2458))$ & 
\footnotesize $2458.0 \pm 1.0 \pm 1.0$ & 
\footnotesize $2456.5 \pm 1.3 \pm 1.3$ & 
\footnotesize $2463.0 \pm 1.7 \pm 1.2$ \\ \hline
\small $\frac{\sigma \cdot \mathcal{B}(D_{sJ}(2458) \rightarrow \! D^{*}_{s}\pi^0)}{\sigma \cdot \mathcal{B}(D^{*}_{sJ}(2317) \rightarrow \! D_{s}\pi^0)}$ & 
\footnotesize $0.25 \pm 0.04$ & 
\footnotesize $0.29 \pm 0.08$ & 
\footnotesize $0.44 \pm 0.13$ \\ \hline 
\strut
\small $\frac{\mathcal{B}(D_{sJ}(2458) \rightarrow \! D_{s}\gamma)}{\mathcal{B}(D_{sJ}(2458) \rightarrow \! D^{*}_{s}\pi^0)}$ & 
--- & 
\footnotesize $0.55 \pm 0.13 \pm 0.08$ & 
\footnotesize $<0.49$ (90\% C.L.) \\ \hline 
\small $\frac{\mathcal{B}(D_{sJ}(2458) \rightarrow \! D^{*}_{sJ}(2317)\gamma)}{\mathcal{B}(D_{sJ}(2458) \rightarrow \! D^{*}_{s}\pi^0)}$ & 
\footnotesize $< 0.22$(95\% C.L.) & --- & $< 0.58$(90\% C.L.) \\ \hline 
\end{tabular}
\label{tab:dsj}
\end{table}

%
\clearpage
\section{Acknowledgements}
%
Many thanks to my collegues in \babar , in particular the charm analysis working group for providing such a stimulating working environment.


%
%
\clearpage


\clearpage

\clearpage

\begin{thebibliography}{99}
\bibitem{cc} The inclusion of charge-conjugate configurations is implicit in this paper.
\vspace{-0.05cm}
\end{thebibliography}
\end{document}